\documentclass[iop,usenatbib]{emulateapj}

\usepackage{epsfig}

\def\HI{\hbox{H\hskip1.5pt$\scriptstyle\rm I\ $}}
\def\HIp{\hbox{H\hskip1.5pt$\scriptstyle\rm I$}}
\def\HII{\hbox{H\hskip1.5pt$\scriptstyle\rm II\ $}}
\def\HIIp{\hbox{H\hskip1.5pt$\scriptstyle\rm II$}}

\def\HeIII{\hbox{He\hskip1.5pt$\scriptstyle\rm III\ $}}

\def\nHI{_{\rm HI}}

\def\nH{_{\rm H}}
\def\nHI{_{\rm HI}}
\def\nHII{_{\rm HII}}
\def\nHe{_{\rm He}}

\def\nHeII{_{\rm HeII}}
\def\nHeIII{_{\rm HeIII}}
\def\I{_{\rm I}}
\def\II{_{\rm II}}
\def\III{_{\rm III}}

\def\se{^{\rm e}}
\def\H{_{\rm H}}
\def\He{_{\rm He}}
\def\HH{H$_2$\ }

\def\nHH{_{\rm H_2}}
\def\ii{_{\rm i}}

\def\ini{_{\rm ini}}

\def\CMB{_{\rm CMB}}

\def\re-ion{_{\rm ion}}

\def\der{{\rm d}}

\def\T{_{\rm T}}

\def\b{}%_{\rm b}}
\def\typ{_{\rm typ}}

\def\lav{\langle}
\def\rav{\rangle}
\def\Jr{_{\rm J}}

\def\CMB{_{\rm CMB}}

\def\lim{_{\rm lim}}

\def\ee{_{\rm e}}
\def\f{_{\rm f}}

\def\H{_{\rm H}}

\def\u1{^{(1)}}

\def\HH{H$_2$\ }

\def\HI{\hbox{H\hskip1.pt$\scriptstyle\rm I$\ }}

\def\HII{\hbox{H\hskip1.pt$\scriptstyle\rm II$\ }}

\def\He{_{\rm He}}

\def\nHH{_{\rm H_2}}
\def\nHII{_{\rm HII}}

\def\y3{$p_{\rm III}$}
\def\yb3{$p_{\rm III}$\ }
\def\e3{$\epsilon_{\rm III}$}
\def\eb3{$\epsilon_{\rm III}$\ }

\def\der{{\rm d}}

\def\Omb{{\Omega_{\rm b}}}

\def\lsim{\lower.5ex\hbox{\ltsima}}
\def\gsim{\lower.5ex\hbox{\gtsima}}

\def\gtsima{$\; \buildrel > \over \sim \;$}
\def\ltsima{$\; \buildrel < \over \sim \;$}

\newcommand{\beq}{\begin{equation}}
\newcommand{\eeq}{\end{equation}}
\newcommand{\beqa}{\begin{eqnarray}}
\newcommand{\eeqa}{\end{eqnarray}}

\def\e{_{\rm e}}

\def\f{_{\rm f}}

\def\H{_{\rm H}}

\def\u1{^{(1)}}

\begin{document}

\shorttitle{An Improved Treatment of IGM Evolution}
\shortauthors{Manrique and Salvador-Sol\'e}

\title{An Improved Treatment of Cosmological Intergalactic Medium
  Evolution}

\author{Alberto Manrique\footnote{E-mail: a.manrique@ub.edu}, and
  Eduard Salvador-Sol\'e}

\affil{Institut de Ci\`encies del Cosmos. Universitat de Barcelona,
  UB-IEEC.\\ Mart\'\i\ i Franqu\`es 1, E-08028 Barcelona, Spain}

\begin{abstract}
  The modeling of galaxy formation and reionization, two central
  issues of modern cosmology, relies on the accurate follow-up of the
  intergalactic medium (IGM). Unfortunately, owing to the complex
  nature of this medium, the differential equations governing its
  ionization state and temperature are only approximate. In this
  paper, we improve these master equations. We derive new expressions
  for the distinct composite inhomogeneous IGM phases, including all
  relevant ionizing/recombining and cooling/heating mechanisms, taking
  into account inflows/outflows into/from halos, and using more
  accurate recombination coefficients. Furthermore, to better compute
  the source functions in the equations we provide an analytic
  procedure for calculating the halo mass function in ionized
  environments, accounting for the bias due to the ionization state of
  their environment. Such an improved treatment of IGM evolution is
  part of a complete realistic model of galaxy formation presented
  elsewhere.
\end{abstract}

\keywords{cosmology: theory --- intergalactic medium --- galaxies ---
  galaxies: formation}

\section{INTRODUCTION}\label{int}

The evolution of galaxies is intertwined with that of the
intergalactic medium (IGM). Mechanical heating of IGM by active
galactic nuclei \citep{BEtal06,Crea06} and radiative heating by X-rays
produced in supernovae \citep{WR78,DS86,C91,WF91,LS91,OH03} together
with ionizing photons emitted by young stars
\citep{I86,R86,S90,MO90,MO92,E92} modify the temperature and
ionization state of the IGM, which in turn alters subsequent galaxy
formation.

The physics involved in the coupled evolution of IGM and luminous
sources is so complex and covers such a wide range of scales that its
treatment involves important approximations. In fact, most studies
focusing on galaxy formation adopt an IGM with fixed adhoc
properties. Only studies of reionization do follow the IGM evolution
in more or less detail.

IGM evolution is described by a couple of differential equations
for its ionization state and temperature with some source functions
provided by a galaxy model. It is in this latter part where most
approximations and simplifying assumptions are made, depending on
the particular approach followed, namely hydrodynamic simulations
\citep{QKE96,WHK97,NS97,CEtal00,WL03,IEtal07,O08,Tr08,B13,SM13a},
numerical and seminumerical simulations \citep{Zhea07,FGea09,
  ZaEtal11,SM13b}, pure analytic models
\citep{Hea96,TW96,DEtal04,FZH04,A12,KG13}, and semianalytic models
\citep{BR92,E92,S94,MD08,Fontea11,WL13}, each with its pros
and cons.

The treatment of the IGM itself, a composite inhomogeneous multiphase
medium, is not fully accurate either. In principle, the problem is
less severe for hydrodynamic simulations than for (semi)numerical and
(semi)analytic models because these equations apply locally, so one
must not worry about the spatially fluctuating properties of
IGM. However, current simulations do not resolve the different ionized
phases \citep{F12}.

A usual procedure (e.g. \citealt{S94,WL03,Bea06,Zhea07}) is to
consider the IGM as having a simple hydrogenic composition and
constant, uniform temperature, equal to the characteristic temperature
of photoionized hydrogenic gas ($\sim 10^4$ K), and to focus on the
evolution of the ionization state through the simple equation derived
by \citet{SG87}. But the IGM temperature is crucial not only for
estimating the minimum galaxy mass but also for computing the
recombination coefficients, so such an approximation also affects the
ionization state of the IGM.

\citet{HG97} derived the first coupled equations for the ionization
state and temperature of the IGM taking into account the dependence of
the latter on hydrogen and helium abundances and local density of the
gas \citep{MM94}. However, these equations only held for the cooling
phase after ionization, and \citet{HH03a} and \citet{HH03} extended
them to include the ionization period.

But the IGM is also multiphasic (\citealt{MEtal00} and references
therein): the neutral, singly, and doubly ionized regions are
separated. \citet{CF05} derived the equations for IGM evolution since
the dark ages taking into account the full composite, inhomogeneous,
and multiphase nature of IGM. However, instead of taking the average
recombination coefficients in each (ionized) phase they use the value
these coefficients would take for the average (approximately
mass-weighted) IGM temperature. On the other hand, they ignored the
mass exchanges between halos and IGM, although about $90$\% of the
initial diffuse gas ends up locked into halos, and the current IGM
metallicity shows that halos also eject substantial amounts of gas
into the medium.

These mass exchanges affect the volume filling factors of the various
ionized species as well as the mean particle kinetic energy, so they
must be taken into account. In principle, this would introduce one
explicit differential equation for each of the varying comoving
densities. But, taking into account their trivial form (i.e. the
variation in each quantity is equal to the corresponding source
function), these variations can be directly included in the usual
master equations. Note that the IGM metallicity determining its mean
molecular weight also changes. However, the mass fraction in metals in
the IGM is so small ($\sim 10^{-2}$ Z$_\odot$ at $z\sim 5$;
e.g. \citealt{SimEtal11} and \citealt{OEtal13}) that these variations
have a negligible effect.

Lastly, the source functions in the IGM master equations were
calculated by averaging the feedback of luminous objects over ionized
regions, assuming an evolving {\it universal} halo mass function
(MF). Yet, as the mass of halos able to trap gas and to form stars
depends on the temperature and ionization state of the surrounding
IGM, the halo MF itself depends on the environment. That is, the MF of
halos lying in ionized or neutral regions differs. This bias,
hereafter referred to as the ionization-bias to distinguish it from
the well-known mass-bias (e.g. \citealt{TEtal10} and references
therein)\footnote{The mass-bias is the dependence on large-scale mean
  density of the abundance of halos with a given mass.} must thus be
corrected for.

The aim of the present paper is to improve the analytical treatment of
IGM evolution by deriving new more accurate master equations for its
ionization state and temperature, and by estimating the halo
ionization-bias necessary to properly compute the source functions in
these equations. Such an improved treatment of IGM can be incorporated
into any given (semi)numerical or (semi)analytic model of galaxy
formation such as the one developed by \citet{MSS15}. The IGM
properties shown throughout the paper to illustrate the effects of the
new treatment have been obtained from that model.

In Sections \ref{filling} and \ref{temperatures}, we derive the new
equations for the IGM ionization state and temperature,
respectively. In Section \ref{halos}, we derive the halo mass
functions that result in neutral and ionized environments. Our results
are discussed and summarized in Section \ref{summ}.

\section{Ionization State Equations}\label{filling}

The structure of IGM is determined by the ionizing radiation from
luminous sources. UV photons with a short mean free path ionize small
regions around these sources. Their less energetic fraction gives rise
to singly ionized hydrogen and helium bubbles, while the less
abundant, more energetic fraction gives rise to doubly ionized helium
subbubbles. Bubbles and subbubbles grow and progressively overlap or
retract and fragment, depending on the intensity of the ionizing flux
is. In any case, the neutral, singly and doubly ionized phases are
kept well separated at any time.

As mentioned, IGM is not only multiphasic but also inhomogeneous. All
IGM properties, such as temperature, baryon density or \HI number
density, are random fields characterized by their respective
probability distribution functions (PDFs). We are here interested in
the time evolution of the IGM properties averaged over different
regions. When these averages refer to the neutral, singly, and doubly
ionized phases, they will be denoted by angular brackets with
subscripts I, II, and III, respectively; when they refer to regions
encompassing one particular chemical species, such as \HII (i.e. all
ionized regions), the subscript will explicitly indicate that chemical
species; and when the average is over the entire IGM, there will be no
subscript. Averages of the product of several (either correlated or
uncorrelated) quantities are for their joint PDF, so they will differ
in general from the product of the averages of the individual
quantities.

The local comoving density of \HII ions, $n\nHII$, at the cosmic time
$t$ satisfies the balance equation
\begin{equation}
\frac{\der n\nHII}{\der t}=\dot N\nHII -\frac{\alpha\nHI(T)}{a^3}
\,n\nHII\, n\ee\,,
\label{bal0}
\end{equation}
where $n\ee$ is the comoving density of free electrons, $\dot N\nHII$
is the local metagalactic emissivity of \HIp-ionizing photons due to
luminous sources and recombinations, including redshifted photons
emitted and not absorbed at higher $z$'s, and the second term on the
right is the recombination rate density to \HIp. Note that the
temperature-dependent recombination coefficient for optically thin
regions, $\alpha\nHI(T)$ (see e.g. \citealt{Me09,FGea09}), is divided
by the cube of the cosmic scale factor $a(t)$ so as to express it in
comoving units.

Taking the average of equation (\ref{bal0}) over the whole IGM, with
the average of the second term on the right decomposed in the sum of
the averages over the different phases I, II, and III, duly weighted
by their respective volume filling factors, $Q\I=1-Q\nHII$,
$Q\II=Q\nHII-Q\nHeIII$, and $Q\III=Q\nHeIII$, with $Q\nHII$ and
$Q\nHeIII$ standing for the \HII and \HeIII volume filling factors,
respectively defined as $\lav n\nHII\rav/\lav n\nH\rav$ and $\lav
n\nHeIII\rav/\lav n\nHe\rav$, gives rise to the rigorous equation
\beq 
\frac{\der \lav n\nHII\rav}{\der t}=\lav \dot N\nHII\rav
-\sum_{\rm i=II}^{\rm III} \left\lav \frac{\alpha\nHI(T)}{a^3}
\,n\nHII\, n\ee\right\rav\ii Q\ii\,.
\label{bal2}
\eeq
Approximating $\alpha\nHI(T)$ in ionized regions by a uniform value
corresponding to the characteristic temperature $T\typ$ of
photoionized hydrogenic gas ($\sim 10^4$ K), and dividing by the
approximately constant value (ignoring inflows and outflows) of the
mean comoving hydrogen density, $\lav n\H\rav$, we arrive at the
following simple equation for the \HII volume filling factor $Q\nHII$
\citep{SG87},
\begin{equation}
\frac{\der Q\nHII}{\der t}= \frac{\lav{\dot N}\nHII\rav}{\lav n\H\rav}
-\frac{\alpha\nHI(T\typ)}{a^3}\,C\nHII\,\lav n\ee\rav\nHII\, Q\nHII\,,
\label{ionrec}
\end{equation}
where $C\nHII\equiv \lav n\nHII^2\rav^{}\nHII/\lav n\nHII\rav\nHII^2$
is the so-called clumping factor. To write equation (\ref{ionrec}), we
have made two approximations: $\lav n\ee\rav^{}\nHII \approx\lav
n\nHII\rav^{}\nHII$ and $\lav n\H\rav^{}\nHII\approx \lav
n\H\rav$. The former presumes hydrogenic composition, and the latter
presumes that ionized regions have the same average properties as the
whole IGM.

However, $\lav n\H\rav$ is not constant, but evolves due to inflows
and outflows into and from halos. In addition, the IGM is not strictly
hydrogenic, as its temperature varies both in space and time. Lastly,
there should be, as mentioned earlier, some halo ionization-bias, so
the average IGM properties in ionized regions should differ in general
from the global average properties. We should thus try to do better.

Let us comeback to the rigorous equation (\ref{bal2}). Neglecting
metals, we have $n\ee=n\nHII(1+Y/4X)+n\nHeIII$, where the comoving
density of \HeIII ions, $n\nHeIII$, takes the approximate form
$f(X,Y,\gamma)n\nHII$, with $f$ equal to a universal function of the
hydrogen and helium mass fractions, $X$ and $Y$, respectively, and the
typical spectral index $\gamma$ of ionizing sources. Thus, the average
in the summation on the right of equation (\ref{bal2}) splits into a
sum of two products of the form: average of a function of $T$ times
average of $n\nHII^2$. This is possible thanks to the fact that there
is essentially no correlation between $T$ and $n\nHII$. The reason for
this is that, in ionized regions, $n\nHII$ is essentially equal to
$n\H=X\,n\b$, where $n\b$ is the baryon density. Furthermore, the only
terms in equation (\ref{T}) for the evolution of the IGM temperature
coupling $n$ and $T$ are the second and fifth ones giving the
heating/cooling by adiabatic compression/expansion of the fluid
element, and the heating/cooling by the loss/gain of baryons due to
inflows/outflows, respectively, which are less than the first term
giving the cosmic adiabatic cooling, and much less than the third and
fourth terms including the stochastic effects of nearby luminous
sources. Under these justified approximations, equation (\ref{bal2})
becomes
\beq
\frac{\der \lav n\nHII\rav}{\der t}=\lav \dot N\nHII\rav - \left\lav \frac{\alpha\nHI(T)}{X\mu\se a^3}\right\rav\nHII 
C\nHII\,\lav n\H\rav^2\,Q\nHII\,,
\label{bal3}
\eeq
where $\mu\se$ is the electronic contribution to the mean
molecular weight. Then, dividing equation (\ref{bal3}) by $\lav
n\H\rav$, we arrive at the new equation
\beqa
\frac{\der Q\nHII}{\der t}=\frac{\lav{\dot N}\nHII\rav}{\lav
  n\H\rav} \nonumber~~~~~~~~~~~~~~~~~~~~~~~~~~~~~~~~~~~~~~~~~~~~~~\\-\left[
\left\lav \frac{\alpha\nHI(T)}{\mu\se a^3}\right\rav\nHII
C\nHII\,\lav n\b\rav
+\frac{\der
  \ln \lav n\H\rav}{\der t}\right]Q\nHII\,.~~~
\label{QHII}
\eeqa
Moreover, taking the Taylor expansion around the average temperature
in phase i, $\lav T\rav\ii$, of the function of temperature $f(T)$
given by the first term in claudators on the right-hand side of
equation (\ref{QHII}), we find that the average over ionized regions,
i=II + III, of $f(T)$ is well-approximated by $f(\lav
T\rav\ii)+(\der^2 f/\der T^2)_{\lav T\rav\ii}{\sigma^2\T}\ii/2$, where
${\sigma\T}\ii$ is the dispersion in temperatures around the mean.

\begin{figure}
\vspace{-0.2cm}
\centerline{\psfig{file=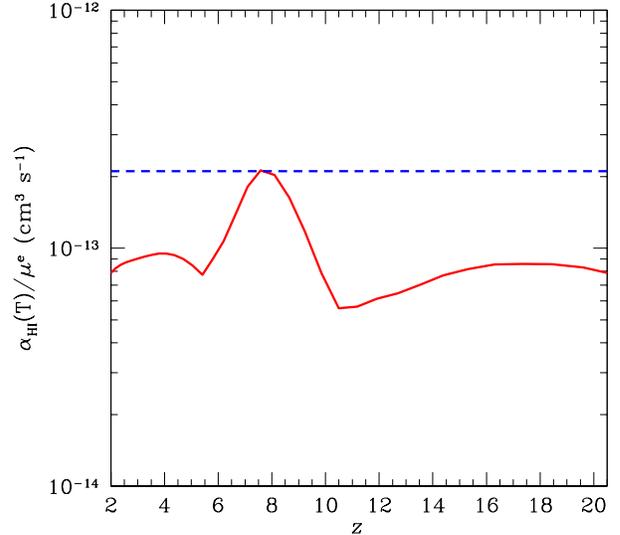,width=.48\textwidth,angle=0}}
\vspace{-0.5cm}
\caption{Average of the recombination to \HI coefficient over the
  electron contribution to the mean molecular weight, $\mu\se$, over
  \HII regions as a function of $z$ for the temperature evolution
  shown in Figure \ref{temp} (solid red line), obtained from the
  galaxy model by \citet{MSS15} for realistic values of the parameters
  leading to double hydrogen reionization \citet{SM14}, compared to
  the usual value for a fixed temperature of $10^4$ K (dashed blue
  line).}
\label{recomb}
\end{figure}

Besides being better justified than expression (\ref{ionrec}),
expression (\ref{QHII}) is also more accurate for the following
reasons: i) instead of taking the recombination rate density at a
fixed typical temperature divided by $\lav \mu\se\rav\nHII$, it uses
the $z$-dependent average of $\alpha\nHI(T)/\mu\se$ in the \HII
region, and ii) the last term on the right accounts for the changing
comoving hydrogen density due to inflows/outflows. In Figure
\ref{recomb}, we compare $\lav \alpha\nHI(T)/\mu\se\rav\nHII$ for the
$z$-dependent temperature shown in Figure \ref{temp} to the uniform
constant value $\alpha\nHI(T\typ)/\lav \mu\se\rav\nHII$ with
$T\typ=10^4$ K appearing in equation (\ref{ionrec}). As can be seen,
the difference is noticeable, particularly around the redshifts $z=$
10.3 and 5.5 of complete ionization in the particular galaxy model
with double reionization considered.

When $Q\nHII$ reaches the value of one and $\lav \dot N\nHII\rav$ is
sufficient to balance recombinations, a period of ionization
equilibrium begins in which $Q\nHII$ stays equal to one. However, if
$\lav \dot N\nHII\rav$ becomes insufficient to keep ionized regions
growing (or stable), a recombination will begin. The constant and
decreasing values of $Q\nHII$ in those two regimes are also governed
by equation (\ref{QHII}), in the former case with $\lav \dot
N\nHII\rav$ replaced by the equilibrium value, with the leftover
metagalactic emissivity eventually used, duly redshifted, to ionize
more hydrogen atoms at lower $z$'s.

A similar derivation leads to the homologous equation for the \HeIII
volume filling factor,
\beqa 
\frac{\der Q\nHeIII}{\der t}=\frac{\lav \dot N\nHeIII\rav}{\lav
  n\He\rav}~~~~~~~~~~~~~~~~~~~~~~~~~~~~~~~~~~~~~~~~~\label{QHeIII}\\ 
-\left[\left\lav\frac{\alpha\nHeII(T)}{\mu\se a^3}\right\rav\III C\nHII\,\lav n\b\rav
  +\!\frac{\der \!\ln \lav n\He\rav}{\der
    t}\right]\!Q\nHeIII\nonumber\,.
\eeqa
Again, if at any point a period of \HeIII ionization equilibrium or
recombination takes place, then $Q\nHeIII$ stays equal to one or
begins to diminish, respectively, according to the same equation
(\ref{QHeIII}).

\section{Temperature Equations}\label{temperatures}

Photo-ionization leads to photo-heating of the different IGM
phases. Other heating mechanisms acting on the IGM are Compton heating
by X-rays and by cosmic microwave background (CMB) photons at very
high-$z$ (after decoupling of baryons from radiation at $z\sim
150$). Such heating is partially balanced by the cooling due to
recombinations and desexcitations, cosmic expansion, Comptonization
from CMB photons at low $z$, and collisional cooling (significant only
in very hot neutral regions, if any). In addition, density
fluctuations suffer gravitational contraction/expansion causing extra
heating/cooling. These are the main mechanisms causing the thermal
evolution of the IGM. Below we mention (in italics) a few additional
mechanisms that are included in the present more accurate treatment
(see also \citealt{HG97} for other possible heating and ionizing
mechanisms, due to decaying or annihilating dark matter, not included
herein).

The local temperature of the IGM evolves according to the differential
equation (e.g. \citealt{CF05})
\beqa
\frac{\der T}{\der\ln(1+z)}= T\Big[2+\frac{2}{3}\frac{\der \ln
    (n/\lav n\rav\ii)}{\der\ln(1+z)}\nonumber~~~~~~\\
+\frac{\der \ln
    \mu}{\der\ln(1+z)}+\frac{\der \ln\varepsilon}{\der\ln(1+z)}
-\frac{\der \ln n}{\der\ln(1+z)}\Big].
\label{T}
\eeqa
The first term in claudators on the right, equal to 2, gives the
cosmological adiabatic cooling of the gas element; the second term
gives its adiabatic heating/cooling by gravitational
compression/expansion for the baryon density $n\b$ around the mean
value $\lav n\b\rav\ii$ in region i, taking into account that most
diffuse IGM is in a linear or moderately non-linear regime; the third
term gives the cooling due to the {\it increase in mean molecular
  weight, $\mu$, caused by ionization and outflows from halos}; the
fourth term gives the Compton cooling from CMB photons, and the
gain/loss of energy density, $\varepsilon$, due to
photo-ionization/recombination, Compton heating from X-rays, {\it the
  achievement of energy equipartition by newly ionized/recombined
  fraction of gas} (the different phases have distinct temperatures in
general) plus {\it mechanical heating accompanying outflows from
  halos}; and the fifth term gives the cooling/heating by {\it the
  gain/loss of baryon density, $n\b$, due to outflows/inflows} (this
changes the average specific energy of the IGM). As outflows take
place from halos harboring luminous sources, we assume that they only
affect ionized regions.

Multiplying equation (\ref{T}) by $\mu \, n$, and taking the average
over each specific phase under the approximation, for the reasons
mentioned in Section \ref{filling}, that $\mu$, $\varepsilon$, $n$,
and $T$ do not correlate with each other, we arrive at
\beq 
\frac{\der\ln \lav T\rav \ii}{\der\ln(1+z)}= 2+\frac{\der \ln(\lav \mu\rav \ii\lav \varepsilon\rav \ii/ \lav n\rav\ii)}{\der\ln(1+z)}\,,
\label{Tio}
\eeq
with i=I, II, or III. Note that, in neutral regions (i=I), there are
no stochastic effects of luminous sources: $\varepsilon$ does not
change either through photo-ionization or by X-rays, $\mu$ is kept
strictly equal to the primordial value, and there is only a small
change in $n$ due to inflows. Consequently, a strong correlation is
foreseen between the quantities $\varepsilon$ and $n$ and
temperature. Yet, we still ignore such a correlation for
simplicity. This approximation is only necessary during the initial
period of increasing ionization; in recombination periods, the gas
properties in the new neutral phase remain uncorrelated as they have
suffered important stochastic feedback effects from luminous objects
over the previous ionized phase.

And what about the temperature dispersion around the mean in the
different IGM phases, also required in equations (\ref{QHII}) and
(\ref{QHeIII})? To calculate ${\sigma^2\T}_i=\lav T^2\rav_i- \lav
T\rav^2_i$ we need to consider the relation
\beq 
\frac{\frac{1}{2}\der T^2}{\der\ln(1+z)}\!=\!
T^2\!\left[2+\frac{2}{3}\frac{\der\ln (n/\lav
    n\rav\ii)}{\der\ln(1+z)}\!+\!\frac{\der \ln
    (\mu\varepsilon/n)}{\der\ln(1+z)}\right]
\label{T2}
\eeq 
following from equation (\ref{T}). The same steps above lead to
\beq
\frac{\der \ln \lav T^2\rav\ii^{1/2}}{\der\ln(1+z)}\!=\! 2+
\frac{\der\ln(\lav \mu\rav \ii\lav \varepsilon\rav \ii/ \lav n\rav\ii)}{\der\ln(1+z)}=\frac{\der \ln \lav T\rav\ii}{\der\ln(1+z)}.
\label{T3}
\eeq

\begin{figure}
\vspace{-0.2cm}
\centerline{\psfig{file=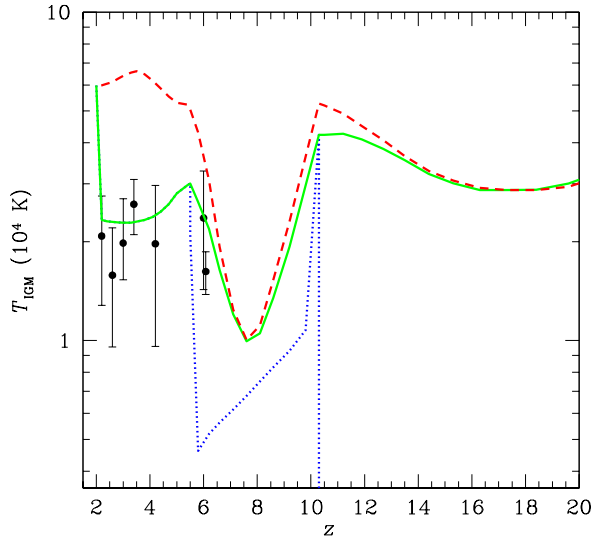,width=.48\textwidth,angle=0}}
\vspace{-.5cm}
\caption{Average IGM temperatures in neutral (blue dotted lines),
  singly ionized (green solid lines), and doubly ionized (red dashed
  lines) regions obtained from the same model with double hydrogen
  reionization (at $z=10.3$ and 5.5) and single helium reionization
  (at $z=2$) as in Figure \ref{recomb}. Solid circles with error
  bars are the actual IGM temperatures estimated by \citet{LiEtal10}
  and \citep{BBWHS10,Bolea12}.}
\label{temp}
\end{figure}

The initial conditions for equation (\ref{Tio}) are $\lav
T\rav\I(z\ini)=T\CMB(z\ini)$ and $\lav T\rav\II(z\ini)=\lav\mu\rav\II
\lav T\rav\I(z\ini)/\lav\mu\rav\I$, where $z\ini$ is the redshift at
which the IGM temperature begins to deviate from the temperature of
CMB photons\footnote{Until that time, the residual density of free
  electrons and ions causes the gas to be thermalized by CMB
  photons.}, satisfying $1+z\ini=100(\Omb h^2/0.0125)^{2/5}$, where
$\Omb$ and $h$ are the baryonic density parameter and the Hubble
parameter scaled to 100 km s$^{-1}$ Mpc$^{-1}$. Similarly, the initial
condition for equation (\ref{T3}) is $\lav
T^2\rav\II(z\ini)=\lav\mu\rav\II^2\lav T^2\rav\I
(z\ini)/\lav\mu\rav\I^2$, where $\lav T^2\rav\I(z\ini)$ is equal to
${\sigma^2\T}\CMB(R\Jr,z\ini)+T\CMB^2(z\ini)$, where
${\sigma^2\T}\CMB(z\ini)= 2T\CMB^2(z\ini)\sigma_0(R\Jr,z\ini)^2/3$ is
the CMB temperature variance at the Jeans scale at recombination,
$R\Jr$, evolved to $z\ini$, with $\sigma_0(R,z)$ the 0-order spectral
moment at the scale $R$ and redshift $z$. The reason for the filtering
at the scale $R\Jr$ is that, at smaller scales, there were no
temperature fluctuations at recombination, and the uniform temperature
on those scales only suffered cosmological adiabatic cooling and the
effects of luminous sources, uncorrelated with $T$. If there is a
period of increasing recombination, the initial mean temperature and
variance in the recombined region are equal (except for
different mean molecular weights) to those in the ionized phase giving
it rise. We have checked that ${\sigma^2\T}(\lav T\rav\ii)$ is always
much less than $\lav T\rav^2\ii$, meaning that the second order Taylor
expansion around $\lav T\rav\ii$ of any arbitrary function $f$ of
temperature is really close to the value $f(T)$.

In Figure \ref{temp}, we show the temperature evolution that results
from the present improved treatment of IGM evolution for the
source functions, $\dot N\nHI$, $\dot N\nHeIII$, $\der \ln \lav
\varepsilon\rav\ii/\der \ln (1+z)$ and $\der \ln \lav n\rav\ii\der \ln
(1+z)$, provided by the same galaxy model as in Figure \ref{recomb}.

\section{Halo Ionization-Bias}\label{halos}

The chance that halos with a given mass $M$ at $t$ will trap gas, and
that the trapped gas will cool either through molecular bands or
atomic lines and form metal-poor or metal-rich stars, respectively,
depends on the temperature and ionization state of the IGM in which
the halos are embedded. Consequently, the halo MF itself must vary
between neutral and ionized environments. Note that, given the
homogeneity of the Universe, these probabilities are not a function of
a specific point. In particular, the probability that a given
arbitrary point lies in a ionized or neutral region is uniform and
equal to $Q\nHII(t)$ and $1-Q\nHII(t)$, respectively.

To calculate the probability that a halo with mass $M$ is located in
an ionized region at the cosmic time $t$, $P_{M}(\HIIp, t)$, we will
first consider the conditional probabilities $P_{M}(\HIIp,t \vert
\HIIp, t\f)$ and $P_{M}(\HIIp,t \vert \HIp,t\f)$ that the halo is in
an ionized region at $t$ given that it was either in an ionized or
neutral region, respectively, at its formation at $t\f$. The former of
these two quantities is simply
\begin{equation}
P_{M}(\HIIp,t \vert \HIIp,t\f)=1-P\nHI(t,t\f)
\label{pionion}\,,
\end{equation}
where $P\nHI(t,t\f)$ is the probability that the halo environment
recombines between $t\f$ and $t$ because of the absence of nearby
sufficiently luminous sources. The latter is given by
\begin{equation}
P_{M}(\HIIp,t \vert \HIp,t\f)=P^\star_{M}(t,t\f)+
P\nHII(t,t\f)\,,
\label{pionneu}
\end{equation}
where $P^\star_{M}(t,t\f)$ is the probability that star formation {\it
  begins} to take place in a halo with $M$ lying in a neutral
environment between $t\f$ and $t$ (we say ``begins'' because newborn
stars soon ionize the medium around the halo), and $P\nHII(t,t\f)$ is
the probability that the halo environment will become ionized in the
same period of time because of the presence of nearby {\it external}
ionizing sources.

\begin{figure}
\vspace{-0.2cm}
\centerline{\psfig{file=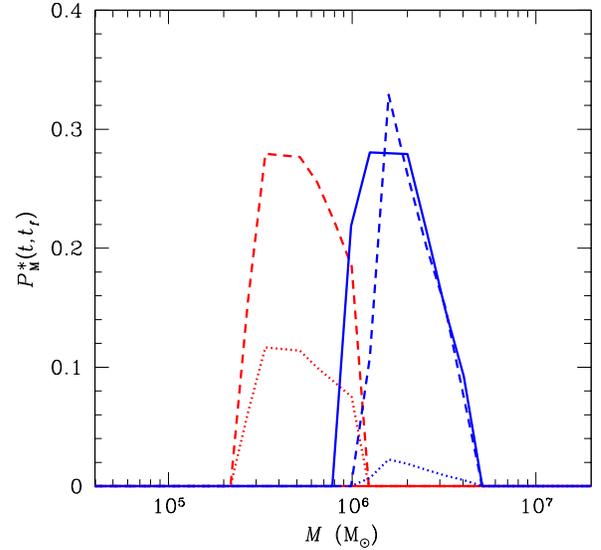,width=.48\textwidth,angle=0}}
\vspace{-0.5cm}
\caption{Probability that star formation will begin to take place in
  halos with $M$ at $z=15$ (blue lines on the right) and 30 (red lines
  on the left), before $100$ Myr (solid curves), $30$ Myr (dashed
  curves), and $10$ Myr (dotted curves) after their formation in
  neutral regions. Note that at $z=30$ there are not yet any halos
  $100$ Myr old.}
\label{probability}
\vskip 5pt
\end{figure}

To derive equations (\ref{pionion}) and (\ref{pionneu}) we have
assumed that the probabilities $P\nHI(t,t\f)$ and $P\nHII(t,t\f)$ are
independent of halo mass. This may not be the case if there is some
correlation between the halo mass- and ionization-biases. However, in
terms of the effect of density on the ionization state of a region the
tendency for halos harboring more powerful ionizing sources to lie in
higher-density regions contrasts with that for ionized bubbles to
stretch more rapidly in lower-density regions, so they tend to balance
one another. Therefore, even though the importance of this correlation
is hard to assess without performing accurate hydrodynamic simulations
with ionizing radiative transfer, we do not expect it to be too
marked. In other words, the present treatment should be reasonably
approximate.

\begin{figure}
\vspace{-0.2cm}
\centerline{\psfig{file=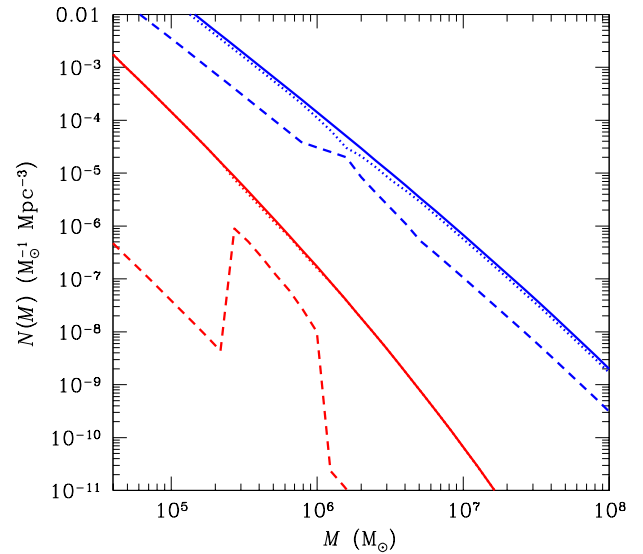,width=.48\textwidth,angle=0}}
\vspace{-0.5cm}
\caption{Halo MF in ionized environments (dashed line) and neutral
  ones (dotted line) at $z=15$ (blue lines on the right) and 30 (red
  lines on the left), compared to the global MF (solid line) at the
  same redshifts, obtained for the same realistic galaxy model as in
  previous Figures. The higher abundance in ionized environments of
  halos in a narrow range of low masses is due to the formation of new
  Population III stars in neutral regions. The rest of the halos in
  ionized regions arise from the ionization the previous objects
  produce around them.}
\label{MF}
\vskip 5pt
\end{figure}

The total probability of finding a halo ionized at $t$ can be
expressed in terms of the above conditional probabilities and
$P_{M}(\HIIp,t)$ upon formation,
%$$
\beqa
P_{M}(\HIIp,t)=P_{M}(\HIIp,t \vert \HIIp,t\f) 
P_{M}(\HIIp,t\f)\nonumber\\
%$$
%\beq
+ P_{M}(\HIIp,t \vert \HIp,t\f)\left[1-P_{M}(\HIIp,t\f)\right]\,.
\label{piontot}
\eeqa 
Substituting the conditional probabilities on the right of
equation (\ref{piontot}) by expressions (\ref{pionion}) and
(\ref{pionneu}), setting $t=t\f+\Delta t$, and taking the limit of
small $\Delta t$, equation (\ref{piontot}) leads to the following
differential equation governing the evolution of $P_{M}(\HIIp,t)$
\begin{equation}
\frac{\der P_{M}(\HIIp,t)}{\der t}=\frac{\der Q\nHII(t)}{\der t}+
\frac{\der P^\star_{M}(t,t\f)}{\der t}\left[1-Q\nHII(t)\right]\,.
\label{difeq}
\end{equation}
To derive equation (\ref{difeq}) we have taken $P\nHI(t,t\f)=0$ and
$P\nHII(t,t\f)=[Q\nHII(t)-Q\nHII(t\f)]/[1-Q\nHII(t\f)]$ in periods of
increasing ionization, and
$P\nHI(t,t\f)=[Q\nHII(t\f)-Q\nHII(t)]/Q\nHII(t\f)$ and
$P\nHII(t,t\f)=0$, in periods of increasing
recombination. Interestingly, in both cases one is led to the same
differential equation (\ref{difeq}), whose solution for the initial
condition $P_{M}(\HIIp,0)=0$ yields the desired probability
$P_{M}(\HIIp,t)$ of finding a halo with $M$ in a ionized region at
$t$, its complementary value giving the probability $P_{M}(\HIp,t)$ of
finding it in a neutral region.

The probability $P^\star_{M}(t,t\f)$ in equation (\ref{difeq})
is hard to estimate analytically because it depends on the number
fraction of \HH molecules, $f\nHH$, at the center of halos with $M$,
whose PDF cannot be established without making appeal to the whole
halo aggregation history. Thus, this function must be drawn from a
full treatment of galaxy and IGM evolution. In Figure
\ref{probability}, we plot this function obtained from the same galaxy
model as in previous Figures. The halo MFs in ionized and neutral
regions resulting from a global MF of the \citet{ST02} form at two
different redshifts are plotted in Figure \ref{MF}. As can be seen,
the higher the redshift, the more marked the effect,which is only 
visible, of course, before full ionization.

\section{SUMMARY}\label{summ}

In the present paper, we have derived an improved version of the
master equations for the evolution of IGM ionization state and
temperature, accounting for the composite, inhomogeneous, multiphase
nature of this medium. Besides all the usual effects, the new version
includes collisional cooling in hot neutral regions (necessary to deal
with recombination periods as found in double reionization), mass
exchanges between halos and IGM, and the achievement equipartition for
newly ionized/recombined gas. In addition, we have derived the
probability that a halo with a given mass $M$ at $z$ is located in a
ionized or neutral environment, which is needed to accurately compute
the source functions required in the IGM master equations.

To check the performance of this improved treatment of IGM we
coupled it to the galaxy model by \citet{MSS15} for realistic values
of the parameters leading to double reionization \citep{SM14}. The
main results were as follows:

\noindent - The average temperatures in the three IGM phases show
marked variations over the different ionization/recombination
periods. This harbors relevant information on the epoch of
reionization. The usual treatment dealing with the average temperature
over the whole IGM (or at mean IGM density, $T_0$) loses this
information.

\noindent - The inclusion of collisional cooling is mandatory to
recover the sudden decrement in the average temperature of neutral
regions after first ionization in double reionization (see
Fig.~\ref{temp}). In the only work to date, by \citet{CF05}, dealing
with the evolution of the average temperature in the different IGM
phases, neutral regions cooled adiabatically after decoupling.

\noindent - The average temperatures of singly and doubly ionized
regions show a maximum similar to that found by Choudhury \& Ferrara
(2005; see panel f of their Fig.~1). However, our temperatures also
show a minimum, due to the recombination after first ionization. More
importantly, the average temperature in doubly ionized regions is
always higher than in singly ionized ones, while this was surprisingly
not the case in Choudhury \& Ferrara's solution.

\noindent - Although the average temperature in singly ionized regions
is not as high as that reported by Choudhury and Ferrara, it is 
still notably higher (by a factor of $\sim 3$) than
the value of $10^4$ K often adopted in reionization studies
\citep{S94,WL03,Bea06,Zhea07}.

\noindent - This difference translates into the average recombination
coefficients. The values we find are substantially smaller (by a
factor $\sim 4$) than found for the temperature of $10^{10}$ K, and
somewhat greater (by a factor $\sim 2$) than the minimum value at the
average temperature reached in Choudhury \& Ferrara's solution.

\noindent - This affects the evolution of the volume filling factors
of ionized hydrogen and helium for identical source functions
(identical galaxy models). But this makes a small difference compared
to that arising from the galaxy models used, which may lead, for
instance, to single or double reionization.

\noindent - We have computed the halo ionization-bias in the
calculation of the source functions appearing in the IGM master
equations. The ratios between the halo MF in ionized and all
environments found for low mass newly star-forming halos and for the
rest are respectively equal to $\sim 3 \times 10^{-4}$ ($0.1$) and
$\sim 0.3$ ($0.4$) at $z=30$ ($z=15$).

This improved treatment of IGM can be easily implemented in any model
of galaxy and IGM evolution. This is particularly advisable for
accurate models of galaxy formation or reionization when contrasting
them with current observations (e.g. \citealt{SM14}) or future ones
(e.g. 21 cm line experiments).

\begin{acknowledgments}
This work was supported by the Spanish DGES grant
AYA2012-39168-C03-02, and the Catalan DIUE grant 2009SGR00217.
\end{acknowledgments}

\end{document}